\documentclass[a4paper]{jpconf}
\usepackage{graphicx}
\usepackage{cite}
\begin{document}
\title{Possible role of gravity in collapse of the wave-function: a brief survey of some ideas}

\author{Tejinder P. Singh}

\address{Tata Institute of Fundamental Research, Homi Bhabha Road, Mumbai 400005, India}

\ead{tpsingh@tifr.res.in}

\begin{abstract}
This article is a brief survey of some approaches to implementing the suggestion that collapse of the wave function is mediated by gravity. These approaches include: a possible connection between the problem of time and problem of quantum measurement, decoherence models based on space-time uncertainty, the Schr\"{o}dinger-Newton equation, attempts to introduce gravity into collapse models such as CSL, ideas based on the black hole - elementary particle complementarity, and the possible role of a complex space-time metric.

\end{abstract}

\section{Introduction}
No experiment to date disagrees with quantum theory. However, it is often not appreciated that the theory has not been tested in all parts of the parameter space (as defined by the number of degrees of freedom in the system being tested).  The largest objects for which the theory has been tested have a mass of about $10^{5}$ amu, and the smallest objects for which classical mechanics is known to hold have a mass of about $10^{-6}$ grams [about $10^{18}$ amu]. This leaves a mesoscopic desert of some thirteen orders of magnitude over which we do not know whether the centre of mass motion of an object obeys quantum mechanics, or classical mechanics, or some new mechanics, to which the classical and quantum are approximations. Nor do we know for sure as to exactly where in this untested parameter space the quantum-classical transition takes place, and what the mechanism for this transition is \cite{Adler:09,Bassi:03,RMP:2012}.

[This is perhaps a suitable place to remark that while macroscopic quantum behaviour is indeed observed in superconductors, superfluids, and Bose-Einstein condensates, this by itself does not preclude the need to test quantum theory in the desert. The macroscopic quantum states in these systems are collective internal states akin to one-particle states, and what we are referring to above is the centre of mass motion of the object as a whole. One should also note that such macroscopic states put very weak bounds on collapse models such as Continuous Spontaneous Localization (CSL) and do not rule out these models \cite{Adler3:07}].

We could ask if we have any reason to suspect that quantum theory might break down in the desert. The quantum measurement problem which has been with us for nearly a century now, does give rise to the belief in a possible breakdown. It might well be that the problem will eventually go away in a suitable re-interpretation of the theory (many-worlds, consistent histories, ...) or in a suitable reformulation (Bohmian mechanics, ...). On the other hand, there remains open the possibility that the theory has to be modified to solve the measurement problem. It is precisely in the untested mesoscopic desert where the modified theories make experimental predictions different from quantum theory, making it all the more relevant that this regime be experimentally tested. If no departures from the standard theory are found, we can be sure that the answer to the foundational problems of quantum theory will lie in a re-interpretation \cite{RMP:2012}.

Mention must also be made of the curious status of the Born probability rule, a rule which has to be invoked in order to explain the random probabilistic outcomes of measurements. The status is curious because even though the Schr\"{o}dinger equation is perfectly deterministic, and the initial conditions unambiguously specified by prescribing the initial wave-function, the interaction of the quantum system with a classical apparatus produces unpredictable outcomes. The only meaningful way to remedy this unphysical situation is to introduce randomness in the initial conditions (Bohmian mechanics) or in the dynamics (collapse models) or to reinterpret measurements in such a way that collapse is only apparent, but not for real. Apart from the need to understand the Born rule, another shortcoming of the standard (Copenhagen) explanation is the vaguely defined classical apparatus - quantum theory has to depend on its own limit for properly understanding outcomes of experiments. This indeed is not satisfactory \cite{Bell,LandauLifshitz}.

In our opinion, the most significant incompleteness in our present understanding of quantum theory is the problem of time, even though this problem gets little attention in the literature. The time which describes evolution in quantum theory is part of a classical space-time whose geometry is produced by classical bodies, which in turn are a limiting case of quantum theory. Once again, the theory has to depend on its own limit for its formulation, and there ought to exist a more complete equivalent formulation which does not refer to classical time. As we briefly note in the next section, maybe when we have such a reformulation we will also understand how to solve the measurement problem \cite{Singh:2012}.

In summary, we may list the four frontiers along which we need to do better with quantum theory: (i) experimental tests of the theory in the yet unexplored part of the parameter space (ii) obtaining a time-independent reformulation of the theory (iii) finding a universally acceptable  resolution of the measurement problem (iv) understanding the origin of the Born probability rule in a deterministic theory.

Models for quantum theory without classical time have not received much attention, and are briefly discussed in the next section. Models which seek to modify quantum theory so as to solve the measurement problem, while taking classical time as given,  are themselves severely constrained on theoretical and experimental grounds. The modified theory should be stochastic nonlinear, and must include anti-Hermitean modifications to the Hamiltonian \cite{Pearle:76,Gisin:81,Gisin:84,Gisin:89,Gisin:95,Weinberg:11}. The most successful and well-studied  such phenomenological model is Continuous Spontaneous Localization (CSL) \cite{Pearle:89, Ghirardi2:90}. CSL bridges the microscopic quantum theory and macroscopic classical mechanics, acting as an underlying generalisation to which the former two are approximations, and a generalisation which solves the measurement problem (collapse is dynamical) and explains the Born rule (strictly speaking, the theory is constructed in such a way that the Born rule follows, and a fundamental explanation of the Born rule still alludes us). 

Since CSL is a phenomenological theory, eventually one would like to look for a deeper explanation for the CSL mechanism, if it proves to be correct. Possible underlying physics could be that the stochastic field which drives the CSL effect is cosmological in origin [maybe having something to do with the cosmic microwave background] or it is a relic thermodynamic effect, as in Adler's theory of Trace Dynamics \cite{Adler:04,Adler:94,Adler-Millard:1996}. Alternatively, the origin of the CSL effect could lie in the universal interaction of gravitation. There are intrinsic quantum fluctuations in the gravitational field, because the objects which produce it are not exactly classical but possess an intrinsic uncertainty, and these fluctuations might feed back to decohere the quantum wave function and possible induce collapse. Such an idea forms the basis for investigating the possible role of gravity in collapse of the wave function., and a possible connection with CSL. Alternatively, maybe it is not the quantum fluctuations of gravity, but some other feature of gravity, not yet appreciated, which might account for collapse.

There is unfortunately no concrete theory to this effect as yet, although there have been a series of investigations. In the subsequent sections of the present modest review, we briefly survey some of the gravity related ideas that have been proposed in the context of quantum collapse. These include: a possible connection between the problem of time and problem of quantum measurement, decoherence models based on space-time uncertainty, the Scr\"{o}dinger-Newton equation and stochastic gravity, attempts to introduce gravity into collapse models such as CSL, ideas based on the black hole - elementary particle complementarity, and the possible role of a complex space-time metric. Earlier reviews of gravity based models include \cite{Diosi:05,Gao,Diosi:2013}.

\section{The problem of time in quantum theory}
The time that is used to describe evolution in quantum theory is classical time. It is a part of classical space-time on which there overlies a space-time geometry which is determined by classical matter fields, in accordance with the laws of general relativity. The Einstein hole argument implies that in order to give operational meaning to points on the space-time manifold, one needs to assign a metric at each space-time point. It evidently follows that classical matter fields are essential for this operational definition, and since such fields are a limiting case of quantum fields, we are forced to conclude that in the definition of the time that quantum theory employs, the theory depends on its own limit (Fig. 1).
\begin{figure} [ht]
  \centerline{\includegraphics{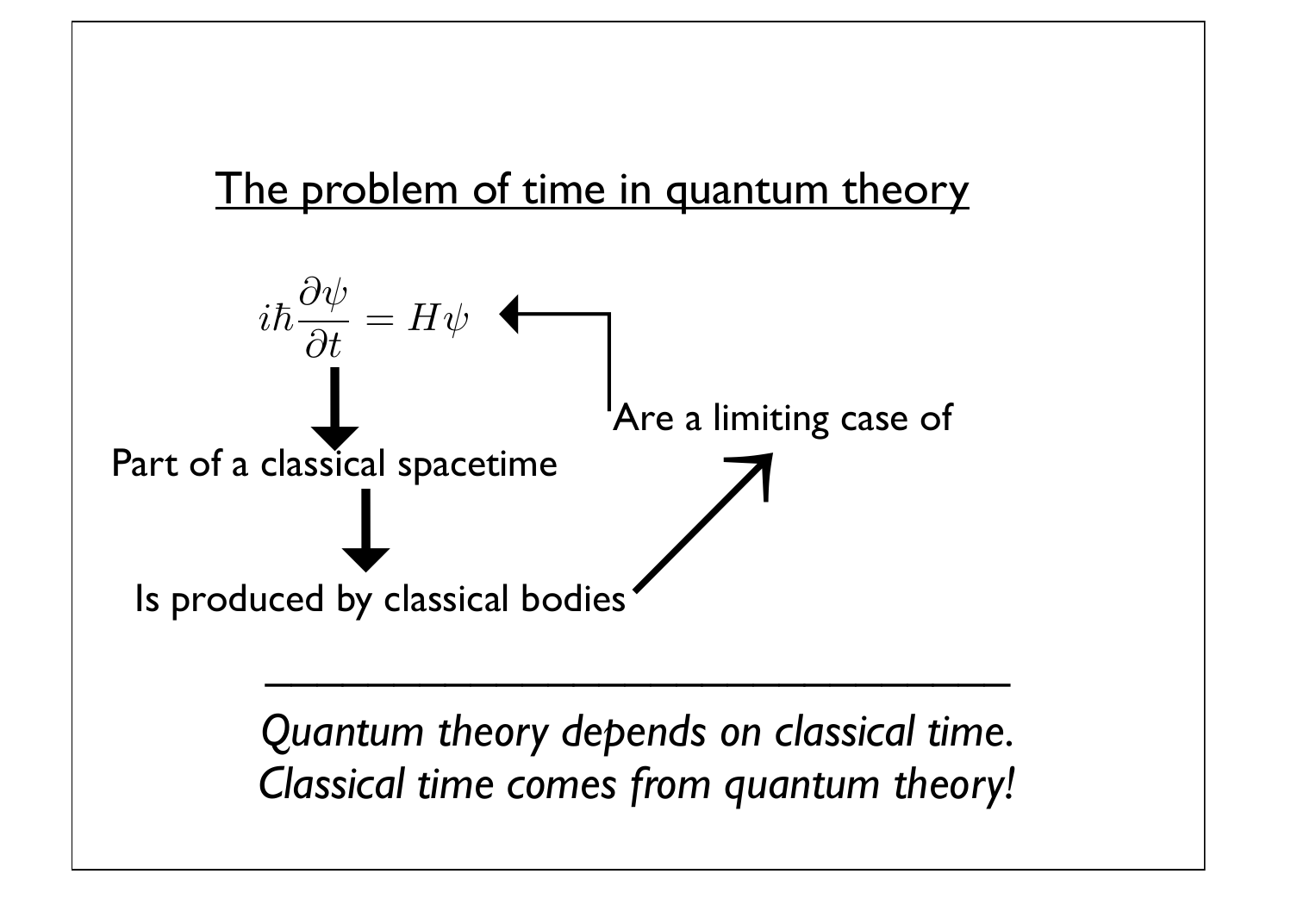}}
\caption{The circularity of time in quantum theory. From \cite{Singh:2012}}  
\end{figure}%

A more complete formulation would be one where such dependence could be avoided, and we call such a formulation `quantum theory without classical time' \cite{Singh:2006}.  For brevity, we venture to name this as pure quantum theory. It is hoped that once such a pure formulation has been developed, standard quantum theory, with its classical time, will emerge as an approximation to the more general formulation. 

Elsewhere we have suggested that pure quantum theory should be formulated by starting from a noncommutative generalisation of special relativity described by the line-element \cite{Lochan-Singh:2011}
\begin{equation}
ds^2 = Tr d\hat{s}^2 \equiv Tr [d\hat{t}^2 - d\hat{x}^2 - d\hat{y}^2 - d\hat{z}^2]
\end{equation}
Space-time coordinates and the matter degrees of freedom that reside on them acquire the status of matrices (equivalently operators). The statistical equilibrium thermodynamics of this noncommutative theory, in parallel with Trace Dynamics, is the sought for pure quantum theory \cite{Lochan:2012}. In the transition from here to a classical world, the role of gravity in causing collapse of the wave-function  is implicated. This is expected to come about because CSL type Brownian fluctuations around equilibrium cause collapse of the wave function describing the quantum space-time in pure quantum theory, as well as of the wave function describing the matter degrees of freedom.

A detailed discussion of this ongoing program can be found in \cite{Singh:2012}, and an overview in the context of collapse theories is given in \cite{RMP:2012}. 

\section{Spacetime uncertainty and decoherence} 
In the absence of a concrete formulation for pure quantum theory, it is useful to build heuristic models by asking how quantum fluctuations in space-time geometry can affect the propagation of the wave-function obeying the Schr\"{o}dinger equation. In order to make progress, one is compelled to start from some sort of description of space-time uncertainty, model it by a classical stochastic potential, and ask how the Schr\"{o}dinger propagation is affected by such a stochastic potential. Decoherence can be expected because of the stochasticity, and it maybe attributed to gravity. One of the earliest such studies was by Karolyhazy, as early as 1966 \cite{Karolyhazi:66} and then followed by Karolyhazy and collaborators in several papers \cite{Karolyhazi:86, Karolyhazy:74, Karolyhazy:90, Karolyhazy:95, Karolyhazy:1982, Frenkel:77, Frenkel:90, Frenkel:95, Frenkel:2002, Frenkel:97}. It was first shown that because of the intrinsic quantum nature of objects which produce gravity, there is a minimal length uncertainty $\Delta s$ in a geodesic of length $\Delta s$, which is given by the following universal relation:
\begin{equation}
\Delta s^3 \sim l_p^2 s
\end{equation}
where $l_p$ is Planck length. 
One may imagine that Minkowski space-time has a perturbing stochastic gravitational field $\gamma({\bf x}, t)$ whose two point correlation is such that it is consistent with the above universal uncertainty. For a matter distribution described by the mass density operator $f({\bf x}, t)$ one then writes the stochastic Schr\"{o}dinger equation:
\begin{equation}
i\hbar\frac{\partial\psi({\bf x}, t)}{\partial t} = \left( H + \int d^3 x' f({\bf x'}) \gamma({\bf x'}, t) \right) \psi({\bf x},t)
\end{equation}

The fluctuating gravity causes distant parts of the wave-function to decohere,  and by considering the evolution of a spherical object of mass $m$ and size $R$ one obtains the following criteria for deciding whether the object is microscopic (quantum dominates over gravity) or macroscopic (gravity dominates over quantum, and the object is classical), and expressions for the critical separation length $a_c$ beyond which decoherence occurs:
\begin{eqnarray}  
a_c\gg R &\rightarrow& \frac{\hbar^2}{G} \gg m^3 R \qquad: micro-region,\\
\nonumber\\
a_c \approx R &\rightarrow& \frac{\hbar^2}{G} \approx m^3 R \qquad: transition-region,\\
\nonumber\\
a_c\ll R &\rightarrow& \frac{\hbar^2}{G} \ll m^3 R \qquad: macro-region.
\end{eqnarray}
There arise two limiting cases: one for $R \gg a_c$ and another for $R \ll a_c$. For a micro-object of linear size $R\ll a_c$, the expression for the coherence length  reduces to that for an elementary particle of mass $m$ and the coherence length $a_c$ over which decoherence effects become relevant is given by
\begin{equation}
a_c \approx \frac{\hbar^2}{G}\; \frac{1}{m^3}  = \left(\frac{L}{l_p}\right)^{2} L; \qquad L= \frac{\hbar}{mc}.
\label{micro}
\end{equation} 
For $R \gg a_c$, the critical length can be expressed as,
\begin{equation}
a_c \approx \left(\frac{\hbar^2}{G}\right)^{1/3}\; \frac{R^{2/3}}{m} = 
\left(\frac{R}{l_p}\right)^{2/3} L.
\label{macro}
\end{equation}

In a different treatment, Di\'osi \cite{Diosi:87, Diosi:07, Diosi:89, Diosi:87a} considered a somewhat different starting point for understanding the intrinsic space-time uncertainty, by asking how much the uncertainty in a measured Newtonian gravitational field would be, if the intrinsic quantum uncertainty of the measuring probe is taken into account. It is shown that if a gravitational field ${\bf g}$ is probed over a volume $V$ for a time interval $T$, then the uncertainty $\delta\tilde{\bf g}$ in the averaged field $\tilde{\bf g}$ is given by
\begin{equation}
(\delta\tilde{{\bf g}})^2 \geq G \hbar/VT\; .
\label{dbound}
\end{equation}
Once again, this is modelled by a classical stochastic perturbing gravitational potential $\phi$ whose two-point correlation is consistent with the above uncertainty bound, and the impact of such a potential on the stochastic Schr\"{o}dinger equation is given by
\begin{equation}
i\hbar\dot{\psi}(t) = \left( H + \int \phi({\bf r}, t) f({\bf r}) d^3r \right) \psi(t)
\end{equation}

The conditions for decoherence and the micro-macro divide are the same as in the Karolyhazy model; however the actual expressions for decoherence length are different:
\begin{eqnarray}
a_{c}&\sim& (\hbar^2/Gm^3)^{1/4}R^{3/4}, \quad {\rm if} \quad Rm^3 \gg \hbar^2/G,\nonumber\\
&\sim& (\hbar^2/Gm^3)^{1/2}R^{1/2}, \quad {\rm if} \quad Rm^3 \ll \hbar^2/G\; .
\label{modelcoh}
\end{eqnarray}
The reasons for this difference were first studied by Di\'osi and Lukacs \cite{Diosi:87a}, and have also been recently studied in \cite{Bera:2014} and are reported by Bera et al. in the present volume.

These models based on modelling of space-time uncertainty can account for decoherence, but do not actually produce collapse to one of the many possible outcomes. This is because the modified Schr\"{o}dinger equation lacks an anti-Hermitean component. This lacking plagues essentially all treatments of collapse models: the anti-Hermitean component has to be put in by hand, for we know what outcome we desire. The fundamental origin for such a component remains to be understood.  

\section{Schr\"odinger-Newton equation and stochastic gravity}
While on the one hand one takes into account the impact on the Schr\"{o}dinger equation of space-time uncertainty produced by other objects, one should also consider how the self-gravity of a quantum object influences its Schr\"{o}dinger evolution. This self-gravity itself has an inbuilt quantum uncertainty and even though we do not really know at present how to describe the self-gravity of a quantum object, we can attempt to model it.  One possible approach is the so-called Schr\"{o}dinger-Newton [SN] equation. It is proposed that that the self-gravitational potential $\Phi$ produced by a quantum source satisfies a semiclassical Poisson equation
\begin{equation}
\nabla^{2}\Phi = 4\pi G m |\psi|^2
\end{equation} 
whose solution is fed into the potential dependent Schr\"{o}dinger equation
\begin{equation}
i\hbar\frac{\partial\psi}{\partial t} = - \frac{\hbar^2}{2m} \nabla^{2}\psi + m\Phi \psi
\end{equation}
so as to obtain the Schr\"{o}dinger-Newton equation \cite{Penrose:96, Penrose:98, Penrose:00, Diosi:84}
\begin{equation}
i\hbar \frac{\partial \psi({\bf r}, t)}{\partial t} =  - \frac{\hbar^2}{2m} \nabla^{2}\psi -
Gm^2 \int \frac{ |\psi ({\bf r'},t)|^2}{|{\bf r}-{\bf r'}|} d{\bf r'} \psi ({\bf r},t)
\end{equation}
This equation has been discussed extensively in the literature \cite{Bernstein:98,giulini2011gravitationally,Harrison:2003,Moroz:98,Ruffini:69,giulini:2012,giulini:2013,Hu:2014,Anastapoulos:2014, Bahrami:2014, Colin:2014}
and also in the present volume by Giulini. The equation is deterministic nonlinear, but not stochastic. This makes it different in character from the Karolyhazy and Di\'osi equation, where by virtue of the stochastic potential the models predict decoherence at the level of the master equation. The SN equation does not predict decoherence but is instead suggested as an equation whose stationary solutions are  ones to which the system will proceed upon decoherence / collapse, once a gravity-based decoherence / collapse mechanism can be incorporated in this system. What has been shown is that there is a gravitationally induced inhibition of dispersion of an expanding wave-packet, at the critical length $a_c \sim \hbar^2 / Gm^3$.

The reason decoherence is not observed is easy to see: one is only dealing with the mean potential in the semiclassical Poisson equation, whereas the Karolyhazy and Di\'osi models incorporate stochastic fluctuations by way of two point correlation of the stochastic potential, which is patterned after the space-time geodesic uncertainty.  One needs to bring in consideration of the stochastic fluctuations, apart from the mean, into the SN equation. One possible way to do this is to employ the theory of stochastic gravity \cite{Hu:2004}, which takes into account corrections to the semiclassical Einstein equations
\begin{equation}
R_{\mu\nu} -\frac{1}{2} g_{\mu\nu} R = \frac{8\pi G}{c^4} \langle\Psi| \hat{T}_{\mu\nu}|\Psi\rangle
\end{equation}
by considering the role of the two-point fluctuations of the energy-momentum tensor [going from semiclassical gravity to the Einstein-Langevin equation].

In order to take into account the effect of stress-tensor fluctuations on gravity, a classical stochastic perturbation is introduced in the metric:
\begin{equation}
g_{\mu\nu} \rightarrow g_{\mu\nu} + h_{\mu\nu}
\end{equation}
and a Gaussian stochastic tensor field $\xi_{\mu\nu}$ is defined though its two-point correlation as follows
\begin{equation}
\langle \xi_{\mu\nu}(x)\xi_{\alpha\beta}(y)\rangle_c = N_{\alpha\beta\mu\nu}
=\langle\{\hat{t}_{\alpha\beta}(x)\hat{t}_{\mu\nu}(y)\}\rangle
\end{equation}
where
\begin{equation}
\hat{t}_{\alpha\beta} = \hat{T}_{\alpha\beta} - \langle\hat{T}_{\alpha\beta}\rangle
\end{equation}  
The Einstein-Langevin equation for the perturbed metric is given by
\begin{equation}
R_{\mu\nu} -\frac{1}{2} g_{\mu\nu} R = \frac{8\pi G}{c^4}
 \left( \langle\Psi| \hat{T}_{\mu\nu}|\Psi\rangle + 2\xi_{\mu\nu} \right) 
\end{equation}
The Newtonian approximation to this equation is given by the modified Poisson equation 
\begin{equation}
\nabla^{2}\Phi = \frac{4\pi G}{c^2} \left ( \langle \Psi | \hat{T}_{00}|\Psi\rangle+ \xi_{00}\right )
\end{equation}
It is hoped that since now the gravitational potential $\Phi$ is stochastic, its incorporation in the SN equation can induce decoherence. This is at present under investigation \cite{Mohan:2015}. A stochastic modification of the Schr\"{o}dinger equation has also been considered by \cite{Nimmrichter:2015}. 

\section{Collapse models and gravity}
A collapse model modifies the Schr\"{o}dinger equation to include a stochastic element, which can dynamically explain the random outcomes of quantum measurements obeying the Born probability rule. The most successful example of a collapse model is [mass-proportional] CSL, described by the equation
\begin{eqnarray} \label{eq:csl-massa}
d\psi_t  =   \left[-\frac{i}{\hbar}Hdt 
 + \frac{\sqrt{\gamma}}{m_{0}}\int d\mathbf{x} (M(\mathbf{x}) - \langle M(\mathbf{x}) \rangle_t)
dW_{t}(\mathbf{x}) \right. \nonumber \\
 -  \left. \frac{\gamma}{2m_{0}^{2}} \int d\mathbf{x}\, d\mathbf{y}\, g({\bf x} - {\bf y})
(M(\mathbf{x}) - \langle M(\mathbf{x}) \rangle_t) (M(\mathbf{y}) - \langle M(\mathbf{y}) \rangle_t)dt\right] \psi_t  
\end{eqnarray}
$H$ is the standard quantum Hamiltonian of the system and the other two
terms induce the collapse of the wave function in space. The mass $m_0$ is a reference mass, which  is usually taken to be nucleon mass. The parameter $\gamma$ is a positive coupling
constant which sets the strength of the collapse process, while $M({\bf x})$ is
a smeared {\it mass density} operator:
\begin{equation}
M(\mathbf{x})
 =  {\sum}m_{j}\delta(\mathbf{x}-\mathbf{r_j}), \label{eq:dsfjdhz}\\
\end{equation}
The smearing function $g({\bf x})$ which encodes the noise correlation is taken equal to
\begin{equation} \label{eq:nnbnm}
g(\mathbf{x}) \; = \; \frac{1}{\left(\sqrt{2\pi}r_{c}\right)^{3}}\;
e^{-\mathbf{x}^{2}/2r_{C}^{2}},
\end{equation}
$r_c$ is the second new phenomenological constant of the model. $W_{t}\left(\mathbf{x}\right)$ is an
ensemble of independent Wiener processes, one for each point in space. Collapse models are discussed in detail in this volume by Bassi.

It would be significant progress if the origin of the CSL mechanism was shown to lie in some universal phenomenon such as gravity. This would imply a gravitational origin for the two new fundamental constants. However, to date such an understanding has not been achieved. In order to incorporate gravity into collapse models Di\'osi \cite{Diosi:89} proposed to replace the CSL noise correlation by the Di\'osi white noise correlation of his model discussed above, and a CSL-like collapse equation
\begin{eqnarray} \label{eq:csl-massa2}
d\psi_t  =   \left[-\frac{i}{\hbar}Hdt 
 + \int d\mathbf{x} (M(\mathbf{x}) - \langle M(\mathbf{x}) \rangle_t)
dW_{t}(\mathbf{x}) \right. \nonumber \\
 -  \left.  \int d\mathbf{x}\, d\mathbf{y}\, g({\bf x} - {\bf y})
(M(\mathbf{x}) - \langle M(\mathbf{x}) \rangle_t) (M(\mathbf{y}) - \langle M(\mathbf{y}) \rangle_t)dt\right] \psi_t  
\end{eqnarray}
The noise correlation is now of gravitational origin, and is given by
\begin{equation}
g({\bf x} - {\bf y}) = \frac{G}{\hbar} \frac{1}{|{\bf x} - {\bf y}|}
\end{equation} 
The corresponding master equation has divergences; to regularise these a cut-off was proposed, at the natural choice of a nucleon scale \cite{Diosi:89}. Unfortunately a cut-off at this scale produces excessive stochastic heating inconsistent with observations. It was then proposed to raise this cut-off to the much higher value $r_c$ which is the new length scale in the CSL model \cite{Ghirardi3:90}. This solves the heating problem but it is difficult to physically justify the inclusion of $r_c$ in the Di\'osi collapse model, one of whose motivations was to achieve a parameter free description of collapse. In an interesting recent discussion \cite{Bahrami:2014a} Bahrami et al. compare the Di\'osi  model with CSL in some detail, and also consider the possibility   of avoiding overheating by introducing dissipation in the dynamics, instead of by going to a high cut-off.

\section{The black-hole elementary particle complementarity}
Considering that black holes and a large class of elementary particles are both fundamental entities characterized  by mass, charge and intrinsic angular momentum, it is suggestive, almost cliched, to consider a possible correspondence between the two.  While the task of searching for such a correspondence is made difficult by the fact that black holes are classical whereas elementary particles are quantum [and an $\hbar$ of course enters in the description of a quantised spin] here we discuss some ongoing work on developing such a correspondence. This work also opens up a possible way in which one might understand involvement of gravity in collapse of the wave-function.

We begin by returning to Eqn. (\ref{macro}), which is Karolyhazy's expression for the coherence length of an extended object. It can be shown that this relation continues to hold for a micro-object as well, and hence it can be treated as a general relation. It can be written in the following useful form
\begin{equation}
\frac{a_c}{R} \approx \left(\frac{L}{R_S}\right)^{2/3}\; \left(\frac{R_S}{R}\right)^{1/3}
\label{ratio}
\end{equation}
where $R_S\sim Gm/c^2$ is of the order of the Schwarzschild radius of the object. If we specialise to the case $R=R_S$ we would be dealing only with black holes [assuming they can be described classically] and this relation becomes
\begin{equation}
\frac{a_c}{R} = \left(\frac{L}{R_S}\right)^{2/3}
\end{equation}

In the macroscopic limit  $a_c\ll R$ the Schwarzschild radius  exceeds the Compton wavelength, and this also implies that $m\gg m_{pl}$, $R_S\gg l_p$ and $L\ll l_p$. The macroscopic black hole limit is obtained for large masses such that Schwarzschild radius is much larger than Planck length, and Compton wavelength is much less than Planck length. Conversely the quantum microscopic limit
$a_c\gg R$ is obtained if  Compton wavelength exceeds Schwarzschild radius, i.e.   $m\ll m_{pl}$, $R_S\ll l_p$ and $L\gg l_p$. The quantum limit is obtained for small masses such that Schwarzschild radius is much less than Planck length, and Compton wavelength is much larger than Planck length: the concept of black hole is not defined in this limit, and we are dealing with a quantum object characterized by its Compton wavelength. It is reassuring that Karolyhazy's analysis for the role of gravity in decoherence produces expected results for the quantum-classical transition from classical black holes to quantum objects.

The macroscopic limit is described accurately by general relativity, accompanied by classical equations of motion for the particle. The 
microscopic limit is described by the Dirac equation  for a relativistic particle of mass $m$. (For now, it is easier to consider the relativistic case, as opposed to the non-relativistic Schr\"{o}dinger equation).  In both these cases [general relativity and the Dirac equation] the mass $m$ acts as a source term. One could enquire as to how the particle `decides' whether it should obey Einstein equations or the Dirac equation? There is no 
information either in Einstein equations or in the Dirac equation which would provide an answer to this question, because neither of the equations involve Planck mass. It is only through experience that we know that micro-objects should be described by the Dirac equation and macro-objects [such as an astrophysical rotating black hole] should be described by the Kerr solution of
Einstein equations. And according to the argument presented above, for gravity induced quantum-classical transition,  the behaviour of an object is quantum or classical, depending on whether or not the associated Compton wavelength exceeds the Schwarzschild radius.  It would be significant if there was a system of equations describing the dynamics of a mass $m$, irrespective of whether or not this mass satisfies $m\ll m_{pl}$ or $m\gg m_{pl}$, and to which system the Einstein equations and the Dirac equation were approximations. Such a system of equations would significantly help in understanding the transition region $m\sim m_{pl}$ as well as provide a more fundamental understanding of gravity induced quantum collapse. 

In order to make some progress towards describing Einstein equations and Dirac equations as part of a common structure, we first look for a common language for describing gravity and the Dirac field. Gravity described via the metric and the Dirac field described by spinors look very different. On the other hand, it is promising to consider Einstein gravity in the tetrad formalism; in particular through the spin coefficients which are employed for writing the Riemann tensor in the Newman-Penrose formalism, via the so-called Ricci identities. The Dirac equation too can be written for a curved space in the Newman-Penrose formalism, and then the set of four Dirac equations look strikingly similar to the Riemann tensor equations [the Ricci identities] written in terms of spin coefficients in the N-P language
\cite{Chandra:43}.

The N-P formalism is a tetrad formalism in which the basis vectors are  null vectors $\bf l, n, m, \overline{m}$ where ${\bf l}$ and ${\bf n}$ are real, and ${\bf m}$ and ${\bf \overline{m}}$ are complex conjugates of each other.  These basis vectors are regarded  as directional derivatives and are denoted by the symbols
\begin{equation}
D={\bf l}, \quad \Delta={\bf n}, \quad \delta = {\bf m}, \quad \delta^{*} = \overline{\bf m}
\end{equation}
Ricci rotation coefficients (i.e. spin coefficients) arise in the definition of the covariant derivatives of these four null vectors. There are twelve such spin coefficients, conventionally denoted by the  standard symbols
\begin{equation}
 \kappa, \sigma, \lambda, \nu, \rho, \mu, \tau, \pi, \epsilon, \gamma, \alpha, \beta
 \end{equation} 

The ten independent components of the Weyl tensor are expressed by five complex Weyl scalars, denoted as
\begin{equation}
 \Psi_0, \Psi_1, \Psi_2, \Psi_3, \Psi_4
 \end{equation}
 The ten components of the Ricci tensor are defined in terms of four real scalars and three complex scalars
 \begin{equation}
 \Phi_{00}, \Phi_{22}, \Phi_{02}, \Phi_{20}, \Phi_{11}, \Phi_{01}, \Phi_{10}, \Lambda, \Phi_{12}, \Phi_{21}
 \label{Ricci}
 \end{equation}
 The Riemann tensor can be expressed in terms of Weyl scalars and Ricci scalars, and directional derivatives of the spin coefficients. There are eighteen complex equations to this effect, known as Ricci identities. A typical Ricci identity has the form
\begin{equation} 
D\rho - \delta^{*}\kappa = (\rho^2+\sigma\sigma^{*}) + \rho( \epsilon + \epsilon^{*})
 -\kappa^{*}\tau -\kappa (3\alpha +\beta^{*}-\pi) + \Phi_{00}
 \end{equation}
The Ricci components are determined from the Einstein equations, and the eighteen complex Ricci identities obey sixteen real constraints known as eliminant conditions, since there are only twenty independent Riemann components \cite{Chandra:43}.

The four Dirac equations for the four spinor components $F_1, F_2, G_1, G_2$ can also be written in the N-P language, and a typical equation has the form
\begin{equation}
(D+\epsilon - \rho) F_1 + (\delta^{*} + \pi - \alpha) F_2 = i\mu_{*} G_{1}
\label{D1}
\end{equation}
where $\mu_* = mc / \sqrt{2}\hbar$. The Dirac equation has a striking similarity with the Ricci identity, with both having a pair of derivatives of spin-coefficients / Dirac spinors.  This similarity can be converted into an actual correspondence, and the four Dirac equations can be recovered as special cases of the Ricci identities, provided eight of the spin coefficients are set to zero
\begin{equation}
\rho=\mu=\tau=\pi=\epsilon=\gamma=\alpha=\beta=0
\label{vanishspin}
\end{equation}
and the following fundamental identification is {\it assumed} between the four Dirac spinors and the remaining four non-zero spin-coefficients:
\begin{equation}
F_1 =\frac{1}{\sqrt{l_p}}\; \lambda, \quad F_2=-\frac{1}{\sqrt{l_p}}\;\sigma, \quad G_1=\frac{1}{\sqrt{l_p}}\;\kappa^{*}, \quad G_2=\frac{1}{\sqrt{l_p}}\;\nu^{*}
\label{match}
\end{equation}
The Dirac equations follow provided one assumes relations between the Riemann components and the Dirac mass, a typical example being
\begin{equation}
\Phi_{20}  + \Phi_{01} = (i\mu_* + \nu)\kappa^*
\label{EDNT1}
\end{equation}
Unfortunately however, the sixteen constraints [the eliminant conditions alluded to above] on the Ricci 
identities lead to undesirable constraints on the Dirac equation, and this particular approach to the Einstein-Dirac correspondence has to be abandoned. 

One then looks for possible circumstances when there are no constraints on the thirty-six real Ricci identities. This can happen when the Riemann tensor has thirty-six independent components, instead of twenty. Indeed there is such a situation possible, and that is when the space-time admits torsion, and hence a Riemann-Cartan geometry, torsion being the anti-symmetric part of the connection:
\begin{equation}
\Gamma_{\mu\nu}^{\ \ \lambda} = \left\{ _{\mu\nu}^{\ \ \lambda} \right\} - K_{\mu\nu}^{\ \ \lambda}
\end{equation}
Here $\left\{ _{\mu\nu}^{\ \ \lambda} \right\}$ is the Christoffel symbol of the second kind, and  $K_{\mu\nu}^{\ \ \lambda}$ is the contortion tensor which describes the presence of torsion.
Now the Ricci tensor is no longer necessarily symmetric, and there are six additional components in the Ricci tensor, described by  the three complex quantities ($\Phi_0, \Phi_1$ and $\Phi_2$).  The Weyl tensor has ten additional components, described by the real quantities ($\Theta_{00}, \Theta_{11}, \Theta_{22}, \chi$) and the complex quantities ($\Theta_{01}, \Theta_{02}, \Theta_{12}$). For definition of these notations the reader is referred to \cite{Jogia}. Thus the Riemann tensor now has exactly thirty-six components, for which there are thirty-six real Ricci identities. There are no eliminant conditions.

The spin coefficients now have an additional term due to torsion, which is denoted by $\gamma_{1}^{nlm}$. Thus, the spin coefficients can now be written as $\gamma_{lmn} = \gamma_{lmn}^{\circ} + K_{nlm} $, where the first term corresponds to the torsion free part and the second term corresponds to the torsion component of the spin coefficients. We further use the following notation to represent the spin-coefficients (see \cite{Jogia} for details):
\begin{equation}
\kappa = \kappa^{\circ} +\kappa_1, \qquad \rho = \rho^{\circ} + \rho_1
\end{equation}
etc.

A few remarks on the fundamental significance of torsion are in order. In special relativity, the symmetry group of the Minkowski space-time is the Poincar\'e group, which includes both Lorentz transformations and translations.  In quantum theory, elementary particles are irreducible, unitary representations of the Poincar\'e group (not the Lorentz group), and these are labelled by mass as well as spin.  On the other hand, in general relativity, which describes a curved space-time, the structure group acting on the tangent space is the Lorentz group, ${\it not}$ the Poincar\'e group. Translations are not included. By introducing torsion and relating it to intrinsic angular momentum, translations are included, and the structure group of the space-time becomes the Poincar\'e group, and is no longer the Lorentz group [Riemann-Cartan geometry]. Curvature is related to Lorentz transformations in the same way that torsion is related to translations \cite{Hehl, Trautmann}. Thus, since both the Dirac equation and Einstein gravity with torsion have the Poincar\'e group as their symmetry group, it seems that correspondence between Einstein equations and Dirac equations might be more easily established if torsion is included. Of course we have to respect the fact that to date there is no observational evidence for torsion in the astrophysical world, and in empty space outside of matter, torsion vanishes. This can be achieved in the construction we now describe.

One first writes down the eighteen complex Ricci identities, now with torsion included in the spin-coefficients \cite{Jogia}. A typical example is
\begin{eqnarray}
%\begin{align}
D\rho - \delta^{*}\kappa = &\ \rho (\rho + \epsilon + \epsilon^{*}) + \sigma \sigma^{*} - \tau \kappa^{*} - \kappa (3\alpha + \beta^{*} - \pi) + \Phi_{00} \nonumber  \\
&- \rho(\rho_1 - \epsilon_1 + \epsilon_{1}^{*} ) - \sigma \sigma_1^{*} + \tau \kappa_1^{*} + \kappa(\alpha_1 + \beta_{1}^{*} - \pi_1 ) + i\Theta_{00}
%\end{align}
\end{eqnarray}
We consider two limits. In the first limit, the torsion part of the spin-coefficients is set to zero. In this limit the Ricci identities reduce to the ones discussed previously, and if the source for the Ricci tensor is taken as the energy-momentum tensor, one recovers Einstein gravity. In the other limit, we set the torsion free part of the spin-coefficients to zero, and retain only the torsion part. We now assume that eight of these torsion dominated spin-coefficients are zero, precisely as in (\ref{vanishspin}) above, and the remaining four non-vanishing spin-coefficients are proportional to the Dirac spinors, as in (\ref{match}).  The Dirac equations then follow from the Ricci identities provided the Riemann tensor obeys the following conditions:
\begin{equation}
\Phi_{20} + i\Theta_{20} + \Phi_{01} + i\Theta_{01} - \Psi_1 - \Phi_0  = i\mu_*\kappa^* 
\label{dir1}
\end{equation}
\begin{equation}
\Phi_{21} + i\Theta_{21} + \Phi_2 - \Psi_3 + \Phi_{02} + i\Theta_{02}  = i\mu_* \nu^* 
\label{dir2}
\end{equation}
\begin{equation}
i\Theta_{12} - \Phi_{12} + i\Theta_{00} - \Phi_{00} + \Phi_{2}^* -\Psi_{3}^* = i\mu_* \sigma 
\label{dir3}
\end{equation}
\begin{equation}
i\Theta_{10} - \Phi_{10} - \Phi_{0}^* - \Psi_{1}^* + i\Theta_{22} - \Phi_{22} = i\mu_* \lambda 
\label{dir4}
\end{equation}
Thus we have the torsion-dominated limit, which are the Dirac equations, and we have the gravity dominated limit, which are the Einstein equations. In the former case, gravity is absent (Minkowski space-time) and the material behaviour is quantum. In the latter case matter behaviour is classical, and gravity dominates. Thus there must be a more general underlying theory in which the torsion-free and the torsion part of the spin-connection are both present, and to which general relativity and quantum theory are both approximations. We conjecture that GR is the $m\gg m_{pl}$ approximation, and Dirac theory is the $m\ll m_{pl}$ approximation, and the more general underlying theory, which is valid for $m\sim m_{pl}$ has the schematic structure
\begin{equation}
\frac{c^3}{G} G_{\mu\nu} - \frac{i\hbar}{l^2} [Riem]^{torsion}_{\mu\nu} = \frac{mc}{l^2} f_{\mu\nu}
\label{genee}
\end{equation}
Although the dynamics of this theory remains to be understood, some general remarks could be made. The length scale $l$ should be determined by the solution of the field equations.
In the first term on the left, $G_{\mu\nu}$ is the Einstein tensor. If this term dominates over the second term, the right hand side  reduces to the symmetric stress-energy tensor. The function 
$f_{\mu\nu}$ then defines an inverse length scale $1/l$, in which case the right hand side
reduces to a mass density times speed of light. It can be inferred that when the second term on the left is negligible, the length scale $l$ is  the Schwarzschild radius.

The second term on the left side is the new idea: the Dirac field could be identified with a complex torsion in a Riemann-Cartan spacetime. If this term dominates over the first term on the left hand side, then by the quantities $[Riem]^{torsion}_{\mu\nu}$ we mean the physics in the left hand side of the four equations (\ref{dir1}) to (\ref{dir4}) above. Now the quantities $f_{\mu\nu}$ on the right hand should be linearly proportional to the corresponding spin coefficients, so that the Dirac equations maybe recovered. If the spin coefficient is of the order $1/l$, then the scale $l$ is found to be of the order of Compton wavelength $L$,  as expected. If the first term dominates we have $R_S\gg L$ and $m\gg m_{pl}$ and if the second term dominates the inequalities are reversed \cite{Sharma2014,Sharma2014a}. 

When $m\sim m_{pl}$ we expect the two terms on the left side to be comparable, and this is the key point of our argument: to propose the Dirac equations as consequence of a modified Einstein gravity, with the modification being caused by inclusion of a complex torsion on a Riemann-Cartan space-time, via the Newman-Penrose equations. We can think of the first term, which contains gravity, as introducing self-gravity into the formalism, which is the modification we seek. Since the non-relativistic limit of the Dirac equation is the Schr\"odinger equation, if we could construct the non-relativistic limit of the gravity-Dirac equation (\ref{genee}) we expect to obtain a non-linear Schr\"odinger equation which incorporates the effect of self-gravity, and which could help understand role of  gravity in collapse of the wave-function.  This work is under progress.

\section{Complex space-time metric, and other ideas}
CSL sets out to construct a model for dynamical collapse based on a stochastic non-linear modification of the Schr\"odinger equation. The structure of the CSL model is uniquely determined by imposing two requirements, whose eventual origin would have to be explained by a more fundamental theory underlying CSL. These requirements are:  preservation of the norm of the state-vector, and no faster than light signalling. An anti-Hermitean linear term is added for obtaining wave-vector reduction, and a supplementary quadratic term is enforced by the requirement of norm preservation. Stochastic noise is introduced to achieve random outcomes of measurements. In order to avoid conflict with experiments \cite{Pearle:94} the noise is assumed to be mass-proportional, and couples to the local mass density operator. This is the preferred form of the CSL model, described in Eqn. (\ref{eq:csl-massa}) above. 
To complete the specification of the model, the form of the noise correlation must be prescribed. 

How does one introduce gravity into the CSL model? One idea, as we saw above, is due to Di\'osi, where the noise correlation of the collapse model is assumed to be the same as in the decoherence model based on gravity. However, this cannot really be called a fundamental approach, because the stochastic part of the Schr\"odinger equation has already been introduced, without ascribing any connection with gravity. In fact, as Adler points out \cite{Adler:2014}, and as we also noted above, real valued metric fluctuations lead to unitary evolution, and can at best induce decoherence, but not collapse of the wave-function. 

An interesting way to relate gravity to actual collapse is to propose that the classical gravitational metric has, apart from the conventional real space-time metric $\overline{g}_{\mu\nu}$, a complex fluctuating term $\phi_{\mu\nu}$, so that \cite{Adler:2014}
\begin{equation}
g_{\mu\nu} = \overline{g}_{\mu\nu} + \phi_{\mu\nu}
\end{equation}
and a noise correlation is prescribed for the fluctuating part. From the standard definition of the energy-momentum tensor in terms of variation of the metric, it can be shown that the imaginary part of $\phi_{00}$ gives rise to a real noise coupling in the CSL equation of the type (\ref{eq:csl-massa}), thus providing a gravitational origin for the CSL noise. It should be noted that the fluctuation term is of an entirely classical nature, and in fact it is assumed that gravity itself does not have to be quantised. Although complex valued classical metrics have been considered earlier in the literature, and in some sense could be considered more natural than real metrics, it remains to understand why the complex part should be stochastic. 

The issue of whether or not gravity should be quantised deserves some more discussion perhaps. If by `quantisation' is meant the application of the rules of quantum field theory to a classical theory of gravity, then the answer is likely to be no. There are various conceptual reasons to believe this, including the one that rules of quantisation are written down assuming a classical space-time background, and it does not appear logically sound to apply those rules to the very background whose pre-existence was assumed for writing down those rules. What then is the status of gravity? Can it be classical, and be described by say semiclassical Einstein equations: a quantum matter source for a classical gravity? Maybe. But then it has been argued that when fluctuations in the quantum energy-momentum tensor operator become significant, the semiclassical description will break down, and will have to be replaced at least by an Einstein-Langevin equation, with the metric acquiring a stochastic component.  To counter such arguments, Dyson has noted that objections against coupling quantum matter to classical gravity apply only to the semiclassical Einstein equations model, and further that the Bohr-Rosenfeld type argument in favour of quantising the electromagnetic field does not apply in the case of gravity \cite{Dyson:2012}. Still, we do not believe that a universe made of only quantum matter fields can coexist 
with a classical space-time manifold: the quantum theory requires a classical time to describe evolution, and classical time is produced by classical bodies: there is no self-consistent demonstration that quantum matter can produce a classical space-time background in which the quantum fields evolve. Perhaps a universe made of only quantum fields requires the space-time to be noncommutative, which in itself is a `nonclassical' feature, which might act as an effective  source of stochasticity and gravity induced collapse. 

We briefly mention three other ideas where gravity has been invoked to explain decoherence or collapse of the wave-function. Pearle and Squires \cite{Pearle:96a} proposed to relate the CSL noise to fluctuations in the Newtonian potential or the curvature scalar.  The theory of Primary State Diffusion due to Percival \cite{Percival:95, Percival:2005}, proposes non-differentiable fluctuations in space-time on the Planck scale to introduce stochastic terms into the evolution equation for a quantum state. The theory bears resemblance to the work of Karolyhazy described earlier. Ellis, Mohanty and Nanopoulos \cite{Ellis:84} discussed how interaction of a microscopic system with space-time wormholes can induce a non-unitary modification in the Hamiltonian evolution. 

Another theoretical testing ground for CSL and gravity induced collapse is the very early universe, where these models have been applied to understand the origin of quantum to classical transition of inflationary density perturbations. While theoretical understanding is still in early stages, many investigations have been carried out, and this promises to be a useful line of research where it might be possible to put constraints on gravity induced collapse from observations of the cosmic microwave background \cite{Das:2014,Das:2013,Martin:2012,Leon:2015,Unanue:08,Castagnino:2014}. Constraints on CSL from the bounds of $\mu-$ and $y-$ distortions of the CMB have been studied in \cite{Lochan:2012a}. 

It has also been suggested that a modification in quantum theory of the CSL type (or gravity induced modification) can help resolve the black hole information loss paradox \cite{Okon:2014}. Thus it seems that time is ripe to consider applications of collapse models in diverse theoretical, astrophysical and cosmological situations, apart from solving the measurement problem or understanding the quantum-classical transition in laboratory systems. Doing so will on the one hand constrain collapse models, and on the other hand might even assist in solving long-standing theoretical puzzles.

\section{Possible experimental tests}   
The most direct experimental tests of departure from quantum theory in the mesoscopic regime are being carried out through molecular interferometry, and optomechanics \cite{Kleckner1999,Marshall:03,RMP:2012,Aspelmeyer:2011,nimmrichter2011testing,Hornberger2011review}. These tests look for a possible breakdown of the quantum superposition principle. No breakdown has been observed to date. Even if a breakdown were to be discovered, it will signal that quantum theory is approximate, but the source of the departure cannot be pinned down to gravity. It is curious that we do not have a gravity specific test, even at the gedanken level, using which we can say that if a violation is discovered, it is necessarily due to gravity. One reason for this is that predictions of gravity induced decoherence / collapse are not sufficiently accurate quantitatively - for instance the quantitative predictions of the Karolyhazy model and the Di\'osi model differ substantially. Theoretical studios have to improve considerably in their precision and reliability, before a gravity specific test can be proposed.  In the same vein, bounds on CSL models coming from laboratory and astrophysical data are also bounds on gravity models \cite{RMP:2012}.

Another promising avenue, apart from interferometry and optomechanics, is a laboratory test for anomalous random walk of a quantum object induced by CSL or by gravity. It is known that the CSL effect induces a tiny stochastic heating of the collapsed particle, leading to a very small energy-momentum violation, too small to contradict any of the known data. However, the momentum violation induces an anomalous random walk which can be in principle be detected at low temperatures and extremely low pressures. Such an experiment has been proposed, but not yet planned or executed. The CSL induced random walk competes with two other standard effects, these latter two being thermal Brownian motion of the particle under consideration, coming from its interaction with ambient radiation (scattering, absorption and emission). The second  standard effect is collision with molecules of the ambient gas (normal Brownian motion). Both these effects can be reduced sufficiently by going to pressures of about
a pico Torr (achievable) and a temperature of 100 Kelvin or lower. It also turns out that the anomalous effect  is easier to detect via the rotational displacement of a disc, as opposed to the translational displacement of a sphere, by levitating the disc, its ideal linear size being about $10^{-5}$ cm  \cite{Bera2015, Collett:2003}. 
Gravity induced collapse / decoherence also predicts such a random walk, though the effect is typically a few orders of magnitude smaller than in the case of CSL \cite{Bera2015, Collett:2003}. 
Another promising approach which has recently gained prominence is investigation of spectral line broadening due to stochastic heating \cite{Bassi:2014a,Bassi:2014b}. While the focus has been on CSL, it would be worthwhile looking into gravity induced effects as well.
It seems particularly relevant and important to look for an experimental effect which could discriminate between a gravity induced collapse and a non-gravitational CSL mechanism. 				  
 
\section{Concluding Remarks}
There are indeed very many ideas which have been tried to investigate if decoherence and collapse of the wave-function can be caused by a gravitational effect. Maybe finding a formulation of quantum theory which does not refer to classical time will throw light on the nature and origin of the quantum-classical transition. The models of Karolyhazy, Di\'osi and Percival, though built on heuristic principles, give very suggestive results. Similarly, including stress tensor fluctuations in the Schr\"{o}dinger-Newton equation indicates gravity induced decoherence. The idea of a complex metric with a stochastic imaginary part is also very attractive, as a neat source for the CSL noise. The black hole - elementary particle complementarity suggests that a gravity theory with torsion bridges general relativity and relativistic quantum mechanics contained in the Dirac equation, and the underlying theory incorporates quantum effects as well as self-gravity. None of the models proposed thus far provide a fundamental explanation for the Born probability rule, nor definitively conclude whether or not special relativity needs to be modified when gravity and quantum mechanics are combined. It is difficult at this stage of our understanding to say whether gravity should be `quantised' in the sense of there being a description of gravity which necessarily has quantum fluctuations as opposed to classical stochastic fluctuations. Nevertheless, diverse investigations of this nature, where the gravity-quantum interface is studied to address unresolved issues in quantum theory, have fascinating conceptual facets, and should be pursued vigorously. These should be accompanied by experimental efforts to ascertain the domain of validity of quantum theory.

\vskip 1 in

{\it ``...I would like to suggest that it is possible that quantum mechanics fails at large distances and for large objects. Now, mind you, I do not say that quantum mechanics does fail at large distances, I only say that it is not inconsistent with what we know. If this failure is connected with gravity, we might speculatively expect this to happen such that $GM^2/\hbar c=1$, or $M$ near $10^{-5}$ gm....If there was some mechanism by which phase evolution had a little bit of smearing in it, so it was not absolutely precise, then our amplitudes would become probabilities for very complex objects. But surely, if the phases did have this built in smearing, there might be some consequences to be associated with this smearing. If one such consequence were to be the existence of gravitateion itself, then there would be no quantum theory of gravitation, which would be a terrifying idea..."}

\bigskip

\rightline{- Feynman (1961)}

\newpage

\noindent {\bf Acknowledgements:} I would like to thank the organisers of DICE2014 for giving me an opportunity to speak at the conference. It is a pleasure to thank my collaborators Angelo Bassi, Shreya Banerjee, Sayantani Bera, Suratna Das, Sandro Donadi, Kinjalk Lochan, Satyabrata Sahu,  Seema Satin, Anushrut Sharma, and Hendrik Ulbricht for interaction and discussions. This work is supported by a grant from the John Templeton Foundation ({\#39530}).

\section*{References}

\bibliography{biblioqmts3}

%\begin{thebibliography}{9}

%\end{thebibliography}

\end{document}